\documentstyle[aps,multicol,epsfig]{revtex}  
  
\begin{document}

\title{Bifurcation analysis of the plane sheet pinch}
\author{\sc J\"org~Schumacher$^1$ and Norbert~Seehafer$^2$}
\address{$^1$Fachbereich Physik, Philipps-Universit\"at Marburg,
D-35032 Marburg, Germany}
\address{$^2$Institut f\"ur Physik,
Universit\"at Potsdam, PF 601553, D-14415 Potsdam, Germany}
\date{December 1999}
\maketitle

\begin{abstract}
A numerical bifurcation analysis of the electrically driven plane sheet pinch is
presented.  The electrical conductivity varies across the sheet such as to allow
instability of the quiescent basic state at some critical Hartmann number.  The
most unstable perturbation is the two-dimensional tearing mode.  Restricting the
whole problem to two spatial dimensions, this mode is followed up to a
time-asymptotic steady state, which proves to be sensitive to three-dimensional
perturbations even close to the point where the primary instability sets in.  A
comprehensive three-dimensional stability analysis of the two-dimensional steady
tearing-mode state is performed by varying parameters of the sheet pinch.  The
instability with respect to three-dimensional perturbations is suppressed by a
sufficiently strong magnetic field in the invariant direction of the
equilibrium.  For a special choice of the system parameters, the unstably
perturbed state is followed up in its nonlinear evolution and is found to
approach a three-dimensional steady state.
\end{abstract}

\begin{multicols}{2}
\section{Introduction}
Finite-resistivity plasma instabilities play an important role for the release
of stored magnetic energy in many astrophysical objects.  They also restrict the
plasma stability in several fusion devices \cite{bis93}.  The simplest
configuration in which they appear is the plane sheet pinch.  In a pinch a
conducting fluid can be held together by the action of an electric current
passing through it with the pressure gradients being balanced by the Lorentz
force.  Of special interest is the resistive tearing instability, which was
studied by Furth {\it et al.}  \cite{furkilros63} by using a boundary layer
approach and afterwards numerically without making the boundary-layer
approximation \cite{wes66,snakil78,stehov83}.  All these studies refer to the
infinite Hartmann number case because of their neglect of the kinematic
viscosity.  The Hartmann number $Ha$, which is the geometric mean of two
Reynolds-like numbers, one being kinetic and the other magnetic, is the
essential parameter that determines the global stability boundaries of the plane
sheet pinch as well as those of its cylindrical counterpart
\cite{dahzanmonhus83,mon92}.  Thus kinematic viscosity has to be included.

A recent sheet pinch study \cite{seeziefeu96} has been done with spatially and
temporally uniform kinematic viscosity and magnetic diffusivity, and with
impenetrable stress-free boundaries.  It is found that the quiescent ground
state (in which the current density is uniform and the magnetic field profile
across the sheet is linear) remains stable, no matter how strong the driving
electric field.  This study was extended to the case of magnetic diffusivity
varying across the sheet, which results in the profiles of the equilibrium
magnetic field deviating from linear behavior.  In particular, the conductivity
profile may be chosen such that the magnetic-field and/or the current profile
have inflection points.  A Squire theorem could be proven for this configuration
\cite{seesch97} whose stability depends on the Hartmann number, the degree of
current concentration about the midplane of the sheet, and on the magnetic
shear (i.e, the asymmetry of the equilibrium magnetic field) \cite{seesch98}.

A stability analysis can be considered as part of a bifurcation analysis, which
will be provided for the cases of two as well as of three spatial dimensions in
the present paper.  In a bifurcation analysis one tries to determine the set of
possible time-asymptotic states, the attractors, for given values of the system
parameters.  The bifurcations from a static sheet-pinch equilibrium have
previously been studied for the case of two spatial dimensions
\cite{sarmas85,sarmas93}.  Grauer \cite{gra89} investigated the interaction of
two different tearing modes in the two-dimensional slab geometry by reducing the
dynamics at the bifurcation point to that on a four-dimensional center manifold.
The new time-asymptotic states were found to be of the tearing-mode type, but
e.g.  also traveling waves were found.

We note here that even though with increasing Hartmann number the equilibrium
becomes first unstable to two-dimensional perturbations (according to the Squire
theorem), the new final states may be three-dimensional.  This is one of the
problems the present paper is addressed to.  The first question arising is which
type of two-dimensional (2D) time-asymptotic state develops nonlinearly from the
tearing mode when the whole problem is restricted to two spatial dimensions.
The bifurcation studies \cite{sarmas85,sarmas93,gra89} predict steady states for
the generic cases.  Similarly, large scale perturbations of a sheared magnetic
field equilibrium were found to result in a final tearing-mode type stationary
state via multiple coalescence of the magnetic island structures
\cite{bofcelpra98}.  A question coming up then is whether the 2D time-asymptotic
states are stable with respect to three-dimensional perturbations.  If not, how
do the stability properties depend on parameters like the strength of a constant
external magnetic field or the wavelength of perturbations in the invariant
direction of the 2D state?  Finally, what are the characteristic properties of
the three-dimensional time-asymptotic states, when they manifest?  In the
present paper for the first time a comprehensive stability analysis of the
two-dimensional time-asymptotic states that develop from the tearing mode is
presented.  For the case of a spatially uniform resistivity these problems were
addressed in numerical studies of the magnetohydrodynamic (MHD) equations by
Dahlburg {\it et al.}  \cite{dahantzan92,dah97}, who found two-dimensional
quasi-equilibria of the tearing-mode type to be unstable to three-dimensional
perturbations.  Secondary three-dimensional instabilities were similarly
observed for non-static primary states with, in addition to a sheared magnetic
field, a pressure-driven jet-like flow \cite{dahkar94}, and their nonlinear
development was proposed as a scenario for the transition to MHD turbulence.

The outline of the paper is as follows.  In Sec.  II the physical model is
introduced.  The MHD equations as well as boundary and initial conditions are
discussed.  In Sec.  III we provide the results of the bifurcation analysis.
First the 2D results are discussed.  After that we investigate the linear
stability of 2D time-asymptotic states with respect to 3D perturbations.  For
appropriate choices of the external parameters the 2D states prove to be 3D
unstable.  The 3D asymptotics is investigated by a full three-dimensional
long-time simulation of the pinch dynamics.  Finally we discuss our results and
end with an outlook in Sec.  IV.

\section{Physical model}
\label{sec_basic}
\subsection{MHD equations}
We use the nonrelativistic, incompressible MHD equations,
\begin{equation}
\label{nseq}
\rho\left(\frac{\partial{\bf v}}{\partial t}+
({\bf v}\cdot{\bf \nabla}){\bf v}\right)
=\rho\nu{\bf \nabla}^2{\bf v}
-{\bf \nabla} p
+ {\bf J}\times{\bf B},
\end{equation}
\begin{equation}
\label{indeq}
\frac{\partial{\bf B}}{\partial t}=
-{\bf \nabla}\times(\eta\mu_0{\bf J}-{\bf v}\times{\bf B}),
\end{equation}
\begin{equation}
\label{diveq}
{\bf \nabla}\cdot{\bf v}=0,\hspace{1em}{\bf \nabla}\cdot{\bf B}=0,
\end{equation}
where ${\bf v}$ is the fluid velocity, ${\bf B}$ the magnetic induction, 
$\mu_0$ the magnetic permeability in a vacuum, 
${\bf J}={\bf \nabla}\times{\bf B}/\mu_0$ the electric current density, 
$\rho$ the mass density, $p$
the pressure, $\nu$ the kinematic viscosity, and $\eta$ the magnetic
diffusivity. While $\rho$ and $\nu$ are assumed
constant, $\eta$ varies spatially:
\begin{equation}
\eta({\bf x})=\eta_0\tilde{\eta}({\bf x}),
\end{equation}
where $\eta_0$ is a dimensional constant and $\tilde{\eta}({\bf x})$ a
dimensionless function of position.

Let the pinch width (sheet thickness) $L\!=\!L_1$ and some yet arbitrary field
strength $B_0$ be used as the  units of length and magnetic induction.
Writing $v_A=B_0/\sqrt{\mu_0\rho}$ for the Alfv\'en velocity corresponding to
$B_0$, we transform to dimensionless quantities. Specifically ${\bf x}$, ${\bf
B}$, ${\bf v}$, $t$, $p$, ${\bf J}$, and ${\bf E}$ are normalized by $L$,
$B_0$, $v_A$, $\tau_A=L/v_A$, $\rho_0 v_A^2$, $B_0/(\mu_0 L)$, and $B_0v_A$,
respectively.  The quantity ${\bf E}$ is the electric field.  Equations
(\ref{nseq}) and (\ref{indeq}) then become
\begin{equation}
\label{nseq_nondim}
\frac{\partial{\bf v}}{\partial t}
=-({\bf v}\cdot{\bf \nabla}){\bf v}+M^{-1}{\bf \nabla}^2{\bf v}
-{\bf \nabla} p
+{\bf J}\times{\bf B},
\end{equation}
\begin{equation}
\label{indeq_nondim}
\frac{\partial{\bf B}}{\partial t}=
-{\bf \nabla}\times(S^{-1}\tilde{\eta}{\bf J}-{\bf v}\times{\bf B}),
\end{equation}
where
\begin{equation}
M=\frac{v_A L}{\nu} \text{ and } S=\frac{v_A L}{\eta_0}
\end{equation}
are Reynolds-like numbers based on the Alfv\'en velocity:
$S$ is the Lundquist number and $M$ its viscous analogue.
The geometric mean of the two Reynolds-like numbers gives the Hartmann number,
\begin{equation}
Ha=\sqrt{M\,S}.
\end{equation}
Finally, the dimensionless Ohm's law becomes
\begin{equation}
\label{Ohm}
S^{-1}\tilde{\eta}{\bf J}={\bf E}+{\bf v}\times{\bf B}.
\end{equation}

\subsection{Boundary conditions and static equilibrium}

We use Cartesian coordinates $x_1$, $x_2$, $x_3$ and consider our magnetofluid
in the slab $0<x_1<1$. $x_1$ is referred to as the cross-sheet coordinate.
In the $x_2$ and $x_3$ directions periodic boundary conditions with periods
$L_2$ and $L_3$, respectively, are used. 

The boundary planes are assumed to be impenetrable and stress-free, i.e.,
\begin{equation}
\label{bc_stress-free}
v_1=\frac{\partial v_2}{\partial x_1}=\frac{\partial v_3}{\partial x_1}
=0 \hspace{1em} \text{at } x_1=0,1.
\end{equation}

The system is driven by an electric field of strength $E^*$ in the $x_3$
direction, which can be prescribed only on the boundary.  We further assume
that there is no magnetic flux through the boundary,
\begin{equation}
\label{B_1}
B_1=0  \hspace{1em} \text{at } x_1=0,1.
\end{equation}

Conditions (\ref{bc_stress-free}) and (\ref{B_1}) imply that the tangential
components of ${\bf v}\times{\bf B}$ on the boundary planes vanish, so that
according to Eq.\ (\ref{Ohm})
\begin{equation}
J_2=0, \hspace{1em} J_3=\frac{E^*S}{\tilde{\eta}_b}  \hspace{1em}
 \text{at } x_1=0,1 ,
\end{equation}
where $\tilde{\eta}_b$ is the value of $\tilde{\eta}$ on the boundaries.
The boundary conditions for the tangential components of ${\bf B}$
then become
\begin{equation}
\label{B_2-B_3}
\frac{\partial B_2}{\partial x_1}=\frac{E^*S}{\tilde{\eta}_b}, \hspace{1em}
\frac{\partial B_3}{\partial x_1}=0
 \hspace{1em} \text{at } x_1=0,1.
\end{equation}
A detailed discussion of these boundary conditions is found 
in Ref.\ \cite{seeziefeu96}.

Any stationary state with the fluid at rest has to satisfy the equations
\begin{eqnarray}
-{\bf \nabla} p+{\bf J}\times{\bf B}&=&{\bf 0},\label{static1} \\
{\bf \nabla}\times(\tilde{\eta}{\bf J})&=&{\bf 0}\label{static2} .
\end{eqnarray}
Equations\ (\ref{static1}), \ (\ref{static2}) and the boundary 
conditions are satisfied by the Harris equilibrium
\begin{eqnarray}
\label{eta_Harris}
\tilde{\eta}&=&\cosh^2[(x_1-0.5)/a],\\
\label{Harris_eq1}
{\bf J}&=&{\bf J}^e\!=\!\left(0,0,\frac{1}{a \tanh(1/2a)
                                    \cosh^2[(x_1\!-\!0.5)/a]}\right)\!,\\
\label{Harris_eq2}
{\bf B}&=&{\bf B}^e\!=\!\left(0,\frac{\tanh[(x_1\!-\!0.5)/a]}{\tanh(1/2a)}+ 
\overline{B_2^e},\overline{B_3^e}\right)\!,
\\
\label{Harris_eq3}
p&=&p^e=-\frac{{{\bf B}^e}^2}{2} \label{pressure-eq},
\end{eqnarray}
where $\overline{B_2^e}$ and $\overline{B_3^e}$ are constants. 
The resistivity given by Eq.~(\ref{eta_Harris}) decreases from the boundary
towards the sheet center where it takes on a minimum value.  This is in
accordance with the expectation that the plasma is hotter within the current
sheet combined with the decrease of the typical Spitzer resistivity with
temperature, i.e. $\tilde{\eta}\sim T^{-3/2}$.  
Unlike other studies \cite{dahantzan92,dah97} where the system is
infinitely extended in the cross-sheet ($x_1$) direction, we do not use the
current sheet half width $a$ as the unit of length.  Instead, we normalize to
the finite distance $L\!=\!L_1$ between the two boundary planes.  
The magnetic
field unit, $B_0$, was chosen in such a way that, in the case of
$\overline{B_2^e}=0$, $|B_2^e|=1$ on the boundary planes.

We  use the notations
\begin{equation}
{\bf b}={\bf B}- {\bf B}^e,
\hspace{1em}
{\bf j}={\bf J}- {\bf J}^e,
\end{equation}
where ${\bf v}$ and ${\bf b}$ are our dynamical variables,  for which
the stress-free boundary conditions are now as follows :
\begin{equation}
\label{bc_stress-free_complete}
v_1=\frac{\partial v_2}{\partial x_1}=\frac{\partial v_3}{\partial x_1}
=b_1=\frac{\partial b_2}{\partial x_1}=\frac{\partial b_3}{\partial x_1}
=0 \hspace{1em} \text{at } x_1=0,1.
\end{equation}
We Fourier expand both vector fields into modes $\sim\!\exp\{i(k_2x_2+k_3x_3)\}$
in the $x_2$ and $x_3$ directions.  In the cross-sheet direction $x_1$ sine and
cosine expansions are used in correspondence with the imposed stress-free
boundary conditions (for more details see \cite{seeziefeu96}).  
Dynamical integrations of
the system are performed in Fourier space by means of a pseudo-spectral method
with 2/3-rule dealiasing.  The grid size for the 3D integrations was taken to be
$32\times 16\times 16$ which was found to be sufficient for our low Hartmann
number studies (see Table 1).  Time integration was performed using a
Runge-Kutta scheme with a variable time step.  Compared to similar calculations
with a spatially uniform resistivity, the simulations were extremely
time-expensive since only very short time steps were possible.  Additionally one
has to keep in mind that our spectral resolution is restricted by the evaluation
of the Jacobian necessary for the linear stability analysis of the
time-asymptotic states.  The used resolution results in the inversion of a
$2836\times 2836$ matrix.

\section{Results}
\label{sec_results}

\subsection{2D time asymptotics}

We started with a determination of the stability boundary for the static
sheet pinch equilibrium.  The Squire theorem allowed a restriction to $x_3$
invariant perturbations (i.e., to perturbations with wave number $k_3\!=\!0$).
Furthermore, due to the invariance of the equilibrium in the $x_2$ direction
stability could be tested for each wave number $k_2$ (or the corresponding
Fourier mode) separately.  First, the system was assumed to be infinitely
extended in the $x_2$ direction.  In this case the wave number $k_2$ of a
perturbation can adopt any real value.  Figure~\ref{fig1} shows the stability
boundary in the Hartmann number--wavelength plane for $\overline{B_2^e}\!=\!0$
and $a\!=\!0.1$ ($\overline{B_2^e}$ and $a$ are fixed to these values throughout
the paper).  Since the equilibrium profile $B_2^e(x_1)$ is symmetric
($\overline{B_2^e}\!=\!0$), the value of $\overline{B_3^e}$ has no influence on
the stability (see Ref.\ \cite{seesch98}).  The unstable region lies to the
right of the boundary curve.  For the spatial resolution used, instability sets
in at $Ha\!=\!{Ha}_c\!=\!64.57$ and $k_2\!=\!{k_2}_c\!=\!2.67$.

In calculations using the full nonlinear equations the aspect ratio $L_2$ (in
the 3D case $L_3$ as well) has to be fixed to a finite value.  We have used
$L_2\!=\!4$ in all nonlinear calculations.  There are some subtleties concerning
the onset of instability and the application of the Squire theorem when $L_2$ is
finite, due to the fact that only a discrete set of $k_2$ values is admitted.
With $L_2\!=\!4$ instability sets in at
$Ha\!=\![{Ha}_c]_{L_2=4}\!=\!66.20784$ and
$k_2\!=\![{k_2}_c]_{L_2=4}\!=\!\pi$, corresponding to a critical wavelength
of $2$.  Unstable 3D modes at Hartmann numbers close to $[{Ha}_c]_{L_2=4}$ are
excluded if the aspect ratio $L_3$ is finite (for more details see Appendix).
Figure~\ref{fig9} shows, for $Ha\!=\!70$, a comparison of the growth rate of
the most unstable 2D mode, which has wavelength 2 in the $x_2$ direction, with
the growth rates of the most unstable 3D mode with the same wavelength in the
$x_2$ direction and different wavelengths in the $x_3$ direction.

When $Ha$ exceeds the critical value $[{Ha}_c]_{L_2=4}$ the tearing mode
grows due to a bifurcation where a pair of identical real eigenvalues becomes
positive.  A superposition of the static equilibrium and the most unstable
eigenvector was taken as the initial state to follow up the nonlinear
development of the tearing mode in two spatial dimensions.

After several hundred Alfv\'{e}n times $\tau_A$ convergence to a stationary
state was observed.  This state is of course linearly stable with respect to
two-dimensional perturbations.  It is also clear that the development to the new
time-asymptotic states is decellerated the closer to the critical value
$[{Ha}_c]_{L_2=4}$ the Hartmann number is taken.  This was indicated first by
the convergence of the maximum eigenvalue to zero.  Namely, due to the marginal
stability with respect to translations in the $x_2$ direction, one eigenvalue of
the time-asymptotic state has to vanish.  For this real eigenvalue $\lambda_0$
we had, for instance, $\lambda_0\simeq -5\cdot 10^{-4}$ for $t\!=\!700$ and
$Ha\!=\!66.21$, $\lambda_0\simeq -5\cdot 10^{-4}$ for $t\!=\!500$ and
$Ha\!=\!66.3$ (and $\lambda_0\simeq -2\cdot 10^{-4}$ for $t\!=\!800$ and
$Ha\!=\!66.3$), but already $\lambda_0\simeq -2\cdot 10^{-6}$ for $t\!=\!500$
and $Ha\!=\!67$ (and $\lambda_0\simeq -10^{-8}$ for $t\!=\!800$ and
$Ha\!=\!67$).  The time development of run 1 became extremely slow.  This run
close to $[{Ha}_c]_{L_2=4}$ was performed in order to make it as sure as
possible that secondary bifurcations close to the primary bifurcation point were
not overlooked.  Since the time-asymptotic solutions for $Ha\!=\!66.21$, $66.3$,
and $67$ are all of the same type, the solutions simulated for $Ha\!=\!66.3$
and $Ha\!=\!67$ are likely to belong to a branch originating in the primary
bifurcation.  In Fig.  \ref{fig2} the time developments of the specific kinetic
energy $E_{kin}\!=\!\frac{1}{2V}\int_V\,{\bf v}^2\mbox{d}\,V$, the specific
magnetic energy $E_{mag}\!=\!\frac{1}{2V}\int_V\,{\bf b}^2\mbox{d}\,V$, and
their sum, the total energy $E_{tot}$, are plotted for runs 2 and 3 (run 3 only
shown in the inset).  Nearly perfect steady states are reached for both Hartmann
numbers at later stages.  For $Ha\!=\!67$ it is practically constant in time
at $t=800$.  The amplitude of the eigenvectors is not determined by the
stability analysis.  For $Ha=67$ thus two energetically different initial
conditions were considered and were found to relax
toward the same asymptotic state,
one from energetically below and the other from energetically above the
asymptotic energy (in the first case, not shown in the figure, the energies increase 
as functions of time and then become almost constant).

In Fig.  \ref{fig3} the new asymptotic 
state is shown for $Ha\!=\!S\!=\!M\!=\!67$.
Field lines of ${\bf B}$, stream lines of ${\bf v}$, and contour lines of the
current density component $J_3$ are drawn. One observes a magnetic island
structure with a chain of $X$ and $O$ points, fluid motion in the form of
convection-like cells or rolls, and a filamentation of the original current
sheet. For $J_3$ only the most inner part of the sheet is shown to highlight
the filamentation despite of the dominant $J_3^e$. Two wavelengths in the $x_2$ direction are seen, corresponding to the
fact that $L_2\!=\!4$ and the critical perturbation has wavelength 2.

\subsection{3D secondary instability of the 2D time-asymptotic states}

Once the 2D time-asymptotic states close to the bifurcation point were
calculated with sufficient accuracy, their linear stability with respect to 3D
perturbations could be investigated.  Though the stability boundary of the
quiescent basic state is determined by the Hartmann number, the bifurcating
states may depend on $S$ and $M$ separately.  We have restricted ourselves,
however, to cases with $Ha\!=\!S\!=\!M$.  The two-dimensional states were
extrapolated to three dimensions by continuing them constantly in the $x_3$
direction, and the stability analysis was performed for the resulting 3D
systems.  Since the equilibria were invariant in the $x_3$ direction, stability
could be tested for each wave number $k_3$ separately.  Taking into account just
one wave number, $k_3\!=\!\mp 2\pi/L_3$, in the $x_3$ direction, the aspect
ratio or pinch height $L_3$ was varied.  We also added constant magnetic shear
components $\overline{B_3^e}$ to the saturated 2D states.  This could be done
since such constant field components do not influence a 2D solution:  The
contribution of $\overline{B_3^e}$ to the Lorentz force ${\bf J}\!\times\!{\bf
B}$ in Eq.\ (\ref{nseq}) vanishes since both ${\bf J}$ and the added magnetic
field are in the $x_3$ direction, and its contribution to the term
$\nabla\!\times\!({\bf v}\!\times\!{\bf B})$ in Eq.\ (\ref{indeq}) vanishes as a
consequence of the incompressibility condition $\nabla\!\cdot\!{\bf v}\!=\!0$
[in the incompressible case one has $\nabla\!\times\!({\bf v}\!\times\!{\bf
B})\!=\!  ({\bf B}\!\cdot\!\nabla){\bf v}\!-\!({\bf v}\!\cdot\!\nabla){\bf B}$].

The motivation for adding a $\overline{B_3^e}$ is that in many
applications externally generated magnetic fields are present in
addition to the self-consistently supported ones. In the solar
atmosphere, for instance, current sheets may form when regions of
obliquely directed magnetic field are brought together and will then
in general have a sheetwise field component. In magnetic fusion devices
like the tokamak toroidal magnetic fields, which correspond to
sheetwise fields in plane geometry, are externally applied to stabilize
the confined plasma.

Results of the stability calculations for $Ha\!=\!M\!=\!S\!=\!66.3$ and
$Ha\!=\!M\!=\!S\!=\!67$ are shown in Fig.\ \ref{fig4} where the maximum real
part of the eigenvalue spectrum is plotted against the varying parameters $L_3$
and $\overline{B_3^e}$.  In the case of $\overline{B_3^e}\!=\!0$ the
two-dimensional saturated states are always unstable, namely to
three-dimensional disturbances with a sufficiently large wavelength in the $x_3$
direction (see upper panel in Fig.\ \ref{fig4}).  

At the stability threshold always two identical real eigenvalues become
positive.  The multiplicity two results from the symmetry of the system with
respect to reflections in the planes $x_3=const.$, due to which the linearly
independent modes with wave numbers $+k_3$ and $-k_3$, respectively, become
simultaneously unstable (the periodic boundary conditions, which allow the
decomposition into Fourier modes, are also needed here) \cite{crakno91}.  The
secondary instability to three-dimensional perturbations is always suppressed by
a sufficiently strong field $\overline{B_3^e}$ (see lower panel in Fig.\
\ref{fig4}).  This is in accordance with the general expectation that a magnetic
field impedes motions with gradients in the direction of the field due to the
tension associated with the lines of force.

The closer to the critical value $[{Ha}_c]_{L_2=4}$ the Hartmann number is,
the larger is the minimum wavelength of the unstable perturbations in the third
dimension (see upper panel in Fig.\ \ref{fig4}).  Now the Squire theorem does
not exclude that 3D perturbations {\em to the quiescent basic state} are
unstable immediately above the critical Hartmann number, provided their
wavelengths $2\pi/k_3$ are sufficiently large (cf.  Appendix).  One might
suspect, therefore, that the unstable 3D perturbations to the 2D
time-asymptotic tearing-mode state are also unstable perturbations with respect
to the basic state at the same Hartmann number.  This is not the case, however:
Consider, for example, the curve for $Ha\!=\!66.3$ in Fig.\ \ref{fig4}
(maximum growth rate over wavelength of the perturbation in the $x_3$
direction).  For $L_3\!=\!2\pi/k_3\!=\!7$ one observes 3D instability of the
2D time-asymptotic state.  Can the quiescent basic state for $Ha\!=\!66.3$ be
unstable to a 3D perturbation with wavelength $7$ in the $x_3$ direction?  The
wave number $k_2$ of such a 3D perturbation can take on the values $n\cdot
2\pi/4$, $n\!=\!1,2,3,\dots$ (since we have chosen the fixed aspect ratio
$L_2\!=\!4$).  Squire's theorem connects the 3D perturbation
to a 2D perturbation with wave number
$\tilde{k_2}\!=\![k_2^2+(2\pi/7)^2]^{1/2}$ which is simultaneously unstable or
stable at the Hartmann number $\tilde{Ha}\!=\!(k_2/\tilde{k_2})66.3$.  With
$k_2\!=\!2\pi/4$, i.e., with the smallest possible $|k_2|$, one finds
$\tilde{Ha}\!=\!57.6$, which is below the critical value ${Ha}_c\!=\!64.57$
(see Fig.\ \ref{fig1}).  That is to say, a 3D mode with $k_2\!=\!2\pi/4$ (and
$k_3\!=\!2\pi/7$) cannot be an unstable perturbation to the quiescent basic
state at $Ha\!=\!66.3$.  For the next possible $k_2$ value, $2\cdot2\pi/4$,
one has $\tilde{Ha}\!=\!63.7$, still below the critical value ${Ha}_c$.  For
all higher $k_2$ values the wavelength $2\pi/\tilde{k_2}$ lies clearly below the
unstable wavelength domain (see Fig.\ \ref{fig1}) for
${Ha}_c\!\leq\!Ha\!\leq\!66.3$.

Figures~\ref{fig5} and \ref{fig6} show an unstable 3D eigenstate to the
time-asymptotic 2D state with $a\!=\!0.1$ and $L_2\!=\!4$ at $Ha\!=\!67$
(which is shown in Fig.\ \ref{fig3}).  The fields are shown in Fig.~\ref{fig5} 
in the
$x_2$-$x_3$ plane to underline qualitatively new structures in the third dimension.
As in Fig.~\ref{fig3}, two wavelengths of the perturbation in $x_2$ are shown.
The 2D equilibrium is mixed with the 3D perturbation in the ratio 50\%
equilibrium to 50\% perturbation.  In the perturbed state velocity and magnetic
field have components in the $x_3$ direction and all structures, including the
current filaments, are modulated in this direction.  

Previous analyses of
secondary instabilities of the sheet pinch \cite{dahantzan92,dah97}, as well as
analyses of similar instabilities in hydrodynamic shear flows \cite{orspat83},
have indicated that these instabilities are ideal, i.e., their growth rates
independent of dissipation.  It appears interesting, therefore, to compare the
growth rates of the secondary instability with those of the primary one at the
same Hartmann numbers.  The growth rate of the most unstable 2D tearing mode
(those with wavelength 2) for $Ha\!=\!66.3$ is $8.8\cdot 10^{-4}$, the
corresponding growth rate for $Ha\!=\!67$ is $7.5\cdot10^{-3}$.  A
comparison with the upper panel in Fig.~\ref{fig4} shows that the secondary
instability grows approximately five times as fast as the primary one.  This
agrees with results of Dahlburg {\it et al.}  \cite{dahantzan92,dah97}.
However, since all our calculations were restricted to $S$ and $R$ values close
to the primary bifurcation point, 
where the growth rates of all
primary or secondary modes go through zero or are still negative,
they do not allow yet a characterization of the secondary mode as
ideal or non-ideal; saturation of the growth rate may occur for larger $S$ and
$R$.

\subsection{3D time asymptotics}
Finally, full three-dimensional simulations were performed to follow the
unstable modes in their nonlinear evolution.  The resistivity gradients made the
simulations again extremely time-expensive.  The calculations were thus
restricted to the case $L_3\!=\!L_2\!=\!4$,
$\overline{B_2^e}\!=\!\overline{B_3^e}=0$, and $Ha\!=\!67$.  The asymptotic
2D state was extrapolated constantly into the third dimension and was mixed with
the most unstable 3D eigenstate giving the initial condition.  The first phase
of the full three-dimensional simulation was done with a lower spectral
resolution, namely $16^3$, up to the time $t_0\simeq 2900$.  
In Fig.~\ref{fig7}, the temporal behavior of
the specific energies after this initial growth phase for the next 900
time units is shown which was
calculated with the highest resolution as given in Table~\ref{Tab1}.  The
energies still oscillate slightly, but with a decreasing amplitude, indicating
convergence to a three-dimensional steady state of the sheet pinch 
configuration.

Furthermore, the solution is characterized by a clear and apparently
time-independent spatial structure.  In Fig.\ \ref{fig8a}
corresponding level surfaces of $|{\bf v}|$ are shown at $t=550$.  We found
the same shape of the level surface of $|{\bf v}|$ at $t=900$.  
Additionally, the same level
surfaces are shown for the 2D time-asymptotic state with otherwise the same
parameters in Fig.\ \ref{fig8}.  The comparison indicates that
there is some relation between the two solutions --- the isosurfaces of $|{\bf
v}|$ in the 3D case are obtained from those in the 2D case by a modulation in
the $x_3$ direction.  This suggests, but does not prove, that the unstable 3D
perturbations to the time-asymptotic 2D state do not drive the system to a
completely different solution existing somewhere in phase space, but that 2D and
3D solutions originate simultaneously in the primary bifurcation of the
quiescent basic state.

\section{Summary and outlook}
\label{sec_summary}

We have numerically studied the primary and secondary
bifurcations of an electrically driven
plane sheet pinch with stress-free boundaries.  The profile of the electrical
conductivity across the sheet was chosen such as to concentrate the electric
current largely about the midplane of the sheet and thus to allow instability of
the quiescent basic state at some critical Hartmann number. 
Our results can be summarized as follows. 

(1) The most unstable
perturbations to the basic state are two-dimensional tearing modes.  If the
whole problem is restricted to two spatial dimensions, also the bifurcating new
time-asymptotic state is of the tearing-mode type, namely, a stationary solution
characterized by a magnetic island structure with a chain of $X$ and $O$ points,
fluid motion in the form of convection-like rolls, and a filamentation of the
original current sheet.  We have calculated this state with precision for the
aspect ratio $L_2\!=\!4$ and Hartmann numbers close to the critical one.  In
contrast to the stability boundary the bifurcating new solutions may depend on
the two Reynolds-like numbers of the problem separately.  We have restricted
ourselves to $M=S=Ha$.

(2) The bifurcating steady (time-asymptotic) two-dimensional 
state was tested for
stability with respect to three-dimensional perturbations.  It proved to be
unstable to 3D perturbations with a sufficiently large wavelength in the third
direction.  At the stability threshold always two identical real eigenvalues
become positive (i.e., there are two purely growing unstable eigenmodes).  We
also added constant external magnetic field components along the invariant
direction to the 2D tearing-mode equilibrium.  If these components are
sufficiently strong, they suppress the secondary instability with respect to
three-dimensional perturbations, which is in accordance with the general
expectation that a magnetic field impedes motions with gradients in the
direction of the field and has been noted before \cite{dah97}.

(3) Full three-dimensional simulations were performed to follow the unstable 3D
modes in their nonlinear evolution. The solution seems to converge to a 3D
steady state.  Although velocity and magnetic field have now components in the
invariant direction of the 2D state and all structures are modulated in this
direction, there is still some resemblance to the 2D tearing mode state.  This
suggests, but does not prove, that the unstable 3D perturbations to the 2D
state do not drive the system to a completely different region in 
phase space. 2D and 3D solutions might originate
simultaneously in the primary bifurcation of the basic equilibrium.

Since our calculations were made very close to the primary bifurcation point, we
suppose that the steady 2D tearing-mode state is unstable from the beginning
and that there is a direct transition of the system from the quiescent basic
state to three-dimensional attractors.  This is true if the 2D state is not
stabilized by an external magnetic field along its invariant direction.
Furthermore, a sufficiently small $L_3$ ensures
stability of the 2D state in a certain Hartmann number interval above the
critical value (since the unstable 3D perturbations, whose wavelength must
exceed some threshold value, are then not admitted).  Finally, the aspect ratio
$L_2$ and the magnetic Prandtl number $Pr_m=\nu/\eta_0=S/M$ might
influence the bifurcation scenario, possibly in such a way that in some
parameter ranges the 2D tearing-mode solution bifurcates stably from the basic
state.

Secondary instabilities that succeed primary two-dimensional ones and that lead
to three-dimensionality have been considered as an important step in the
transition from laminar to turbulent states in linearly unstable nonconducting
shear flows \cite{orspat83,bayorsher88,her88}.  For the case of linearly stable
shear flows (e.g.  the plane Couette flow), it was suggested that nonlinear
stationary and linearly unstable 
three-dimensional states, which develop already below the onset
threshold of turbulence \cite{Nag90,Cle97}, can form a chaotic
repellor in phase space \cite{Sch97}. Such a repellor can cause
the transient turbulent states above the onset threshold.
There is some analogy of the magnetohydrodynamic pinch to shear flows, and
Dahlburg {\it et al.}  \cite{dahantzan92,dah97} have presented numerical
evidence that the secondary instability of two-dimensional quasi-equilibria of
the tearing-mode type can lead to turbulence in a plane sheet pinch.  In their
calculations the pinch was not driven by an external
electric field (nor mechanically driven) and the electrical conductivity was
assumed to be spatially uniform.  In such a case the pinch always decays
resistively, that is, velocity and magnetic field tend to zero as $t\to\infty$.
By our choice of the resistivity profile and the applied electric field we could
calculate exact time-asymptotic, in particular steady states and could
corroborate the result of Dahlburg {\it et al.}  that saturated two-dimensional
tearing-mode states are unstable to three-dimensional perturbations.  We did not
observe a transition to a turbulence-like state yet.  Irregular behavior may be
expected to arise through subsequent bifurcations when the Reynolds-like numbers
are further raised.

\acknowledgments J.S. wishes to acknowledge many fruitful discussions with
Armin Schmiegel. We thank the referee for helpful comments.

\appendix
\section*{Instability of the quiescent basic state and Squire's theorem
in the case of a finite aspect ratio $L_2$}
Squire's theorem states that for increasing Hartmann number two-dimensional
perturbations to the quiescent basic state become unstable first.  Specifically:
For each three-dimensional eigenmode with wave numbers $k_2$, $k_3$ and growth
rate $\lambda$ at Hartmann number $Ha$, there exists a two-dimensional (i.e.,
$x_3$ invariant) eigenmode with wave number
$\tilde{k_2}\!=\!(k_2^2+k_3^2)^{1/2}$ and growth rate
$\tilde{\lambda}\!=\!(\tilde{k_2}/k_2)\lambda$ at Hartmann number
$\tilde{Ha}\!=\!(k_2/\tilde{k_2})Ha$ \cite{seesch97}.

\subsection{Case $L_2\!=\!\infty$}
If the stability problem is considered on the infinite $x_2$-$x_3$ plane, i.e.,
with all wave numbers $k_2$ and $k_3$ allowed, for increasing $Ha$ one or
several
two-dimensional modes with a critical wave number ${k_2}_c$ (and the 
corresponding
critical wavelength ${L_2}_c=2\pi/{k_2}_c$)
become unstable at a critical Hartmann number ${Ha}_c$
where all three-dimensional modes are still stable.  Above ${Ha}_c$ the
critical value ${k_2}_c$ broadens to an unstable $k_2$ interval.  The latter
means, however, that three-dimensional modes could be unstable immediately above
${Ha}_c$.  Namely, consider a 3D eigenmode with wave numbers $k_2$, $k_3$ and
growth rate $\lambda$ at some Hartmann number $Ha\!=\!{Ha}_c+\epsilon$,
$\epsilon>0$.  If $k_2$ is chosen from the interior of the unstable $k_2$
interval at $Ha$, then $\tilde{k_2}\!=\!(k_2^2+k_3^2)^{1/2}$ lies within the
unstable $k_2$ interval at the Hartmann number
$\tilde{Ha}\!=\!(k_2/\tilde{k_2})Ha$, where ${Ha}_c < \tilde{Ha} <
{Ha}_c + \epsilon$, if only $|k_3|$ is chosen sufficiently small.  This does
not mean yet that the 2D mode to which the 3D mode is connected is unstable,
since there are in general also stable 2D eigenmodes with the same wave number
$\tilde{k_2}$.  But if the associated 2D mode is unstable, i.e., if the real
part of $\tilde{\lambda}\!=\!(\tilde{k_2}/k_2)\lambda$ is positive, this implies
that also $\Re(\lambda)\!>\!0$.  The possibility of unstable three-dimensional
eigenmodes close to the critical Hartmann number is excluded by the Squire
theorem, however, if $L_3$ is finite, i.e., if there is a
positive lower bound (however small) to the modulus of the wave number $k_3$.
In that case there is a finite Hartmann number interval above ${Ha}_c$ where
all unstable eigensolutions are purely two-dimensional.

\subsection{Case $L_2$ finite}
Fixing $L_2$ to a finite value complicates the problem, since
only a set of discrete values is admitted for $k_2$.  If not just
$L_2\!=\!n\cdot 2\pi/{k_2}_c$, with $n$ denoting a positive integer number, that
is, if ${k_2}_c$ is not just an admissible $k_2$, instability to 2D modes will
set in at some Hartmann number $[{Ha}_c]_{L_2}$ above ${Ha}_c$ and for a
wave number $[{k_2}_c]_{L_2}$ different from ${k_2}_c$.

\subsubsection{Subcase $L_2\leq{L_2}_c=2\pi/{k_2}_c$}
\label{s2a}
In the case $L_2\leq{L_2}_c$ for all $k_2$ holds $k_2\ge{k_2}_c$ and
consequently the smallest admissible $k_2$ becomes unstable first, i.e.,
$[{k_2}_c]_{L_2}\!=\!2\pi/L_2\geq{k_2}_c$.  It is easily seen that as in the
case of $L_2\!=\!\infty$ (i) directly at the onset of instability only 2D modes
can be unstable (since modes with $k_2\!=\!0$ cannot be unstable \cite{seesch98}
and the Squire theorem thus would connect any unstable 3D mode to a 2D mode with
wave number $\tilde{k_2}>[{k_2}_c]_{L_2}$ outside the unstable $k_2$ interval at
the Hartmann number $[{Ha}_c]_{L_2}$), (ii) immediately above $[{Ha}_c]_{L_2}$
also unstable 3D modes are possible (or at least not forbidden by Squire's
theorem), (iii) a finite aspect ratio $L_3$ (however large) ensures that in a
finite Hartmann number interval close to the onset of instability only purely
two-dimensional eigenmodes are unstable (see also Fig.~\ref{fig9} where the pinch
is stable with respect to 3D modes for $L_3\lesssim 6$).

\subsubsection{Subcase $L_2>{L_2}_c=2\pi/{k_2}_c$}
More involved is the situation for $L_2>{L_2}_c$.  Then it cannot be
excluded generally that 3D modes become unstable first, and in principal each
individual situation has to be tested separately.  One can distinguish between
the cases $[{k_2}_c]_{L_2}>{k_2}_c$ and $[{k_2}_c]_{L_2}<{k_2}_c$, of which the
first one is simpler.  In both cases special complications arise from the fact
that 3D modes with wave numbers $k_2$ smaller than $[{k_2}_c]_{L_2}$, that is to
say, with $k_2\!=\!n\cdot2\pi/L_2< [{k_2}_c]_{L_2}\!=\!n_0\cdot2\pi/L_2$ ($n$,
$n_0$ denoting integer numbers) can come into play.  

In the case of $[{k_2}_c]_{L_2}>{k_2}_c$ these 3D modes (with $k_2$ smaller than
$[{k_2}_c]_{L_2}$) are the only 3D modes that could become unstable at a
Hartmann number less than $[{Ha}_c]_{L_2}$ (where the first 2D mode becomes
unstable); if they remained stable, the situation is similar to
subcase~\ref{s2a}.  The 3D modes with $k_2<{k_2}_c$ must remain stable close to
the onset of 2D instability, however, if $[{Ha}_c]_{L_2}$ does not exceed
${Ha}_c$ too much (and $[{k_2}_c]_{L_2}$ does not differ too much from
${k_2}_c$), such that (i) $|k_3|$ has to be larger than some positive threshold
value in order that $\tilde{k_2}\!=\!(k_2^2+k_3^2)^{1/2}$ (with
$k_2\!=\!n\cdot2\pi/L_2$, $n<n_0$) can come into the unstable $k_2$ interval
close to the onset of instability (since there is a finite gap between the
unstable $k_2$ interval and the largest admissible $k_2$ that is smaller than
${k_2}_c$) and (ii) as a consequence of this
$\tilde{Ha}\!=\!(k_2/\tilde{k_2})Ha$ must be smaller than ${Ha}_c$.  If this is
the case and, furthermore, $[{k_2}_c]_{L_2}>{k_2}_c$, the situation is the same
as for $L_2\leq 2\pi/{k_2}_c$.

The numerical example of this paper belongs to the category just discussed:
$L_2$ finite, $L_2>2\pi/{k_2}_c$, $[{k_2}_c]_{L_2}>{k_2}_c$, and close to the
onset of instability no unstable 3D modes with $k_2<[{k_2}_c]_{L_2}$.  
We found ${k_2}_c=2.67$,
corresponding to a critical wavelength of ${L_2}_c=2.35$, and 
${Ha}_c=64.57$.  The
critical values for the fixed aspect ratio $L_2=4$ are
$[{k_2}_c]_{L_2=4}=\pi$, corresponding to a critical wavelength of 
$[{L_2}_c]_{L_2=4}=2$, and
$[{Ha}_c]_{L_2=4}=66.20784$ (see also Fig. \ref{fig1}). Loosely speaking, an
unstable 3D mode has to fit now between $Ha_c$ and
$[{Ha}_c]_{L_2=4}$ with its critical Hartmann number $\tilde{Ha}$. 
It can only have the wave number
$k_2\!=\!2\pi/4\!=\!\pi/2$, since otherwise
$\tilde{k_2}\!=\!(k_2^2+k_3^2)^{1/2}>[{k_2}_c]_{L_2}$ (i.e., the associated 2D
mode would be stable).  
This implies, in order to have instability
\begin{equation}
\tilde{k_2}\!=\![(\pi/2)^2+k_3^2]^{1/2}>{k_2}_c\!=\!2.67\;,
\end{equation}
and consequently
\begin{equation}
\tilde{Ha}\!=\!\frac{\pi/2}{\tilde{k_2}}[Ha_c]_{L_2=4}
<\frac{\pi/2}{{k_2}_c}[Ha_c]_{L_2=4}\!\simeq\!38.9\;,
\end{equation}
which lies below ${Ha}_c$.  3D modes with $k_2\!=\!\pi/2$ cannot be unstable even for
Hartmann numbers significantly above $[{Ha}_c]_{L_2=4}$.  3D modes with
$k_2\!=\![{k_2}_c]_{L_2=4}$, on the other hand, can be unstable immediately
above $[{Ha}_c]_{L_2=4}$ and are stabilized by an upper bound to the aspect
ratio $L_3$, as discussed in the preceding subsections of this Appendix.  We
found the condition $L_3<1000$ to be sufficient to stabilize all 3D modes at
$Ha\!=\!66.208$ ($>[{Ha}_c]_{L_2=4}\!=\!66.20784$).


\begin{narrowtext}
\begin{table}
\caption{Parameters of the 2D and 3D sheet pinch simulations.} 
\begin{tabular}{ccccc}
Run&   $Ha$ & $L_2$ & $L_3$ & $(N_1,N_2,N_3)$\\
\hline
$1$  & 66.21  & 4 & --   & (32,16,--)    \\ 
$2$  & 66.30  & 4 & --   & (32,16,--)    \\ 
$3$  & 67.00  & 4 & --   & (32,16,--)    \\ 
$4$  & 67.00  & 4 & 4    & (32,16,16) \\
\end{tabular}
\label{Tab1}
\end{table}

\begin{figure}
\begin{center}
\epsfig{file=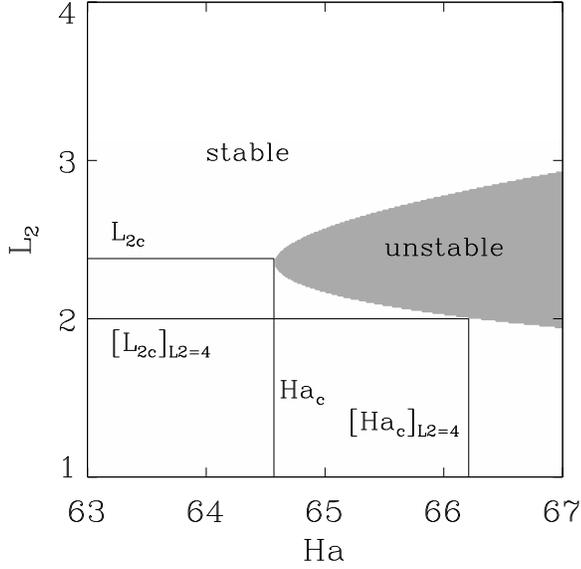,width=7cm,height=7cm}
\end{center}
\caption{Stability boundary for the quiescent basic state with $a\!=\!0.1$ and
$\overline{B_2^e}\!=\!0$ in the $Ha$-$L_2$ plane. $L_2\!=\!2\pi/k_2$ is the
wavelength of the perturbation in the $x_2$ direction (stability was tested
for each wave number $k_2$ separately).}
\label{fig1}
\end{figure}
\begin{figure}
\begin{center}
\epsfig{file=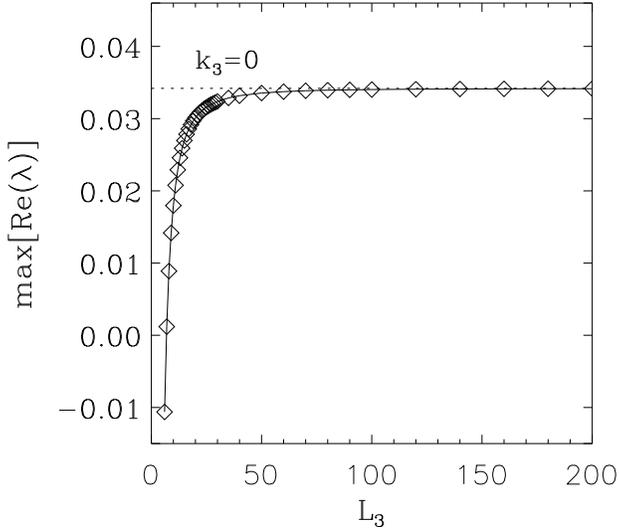,width=6.5cm,height=6.5cm}
\end{center}
\caption{Maximum real parts of the eigenvalue spectrum vs.
$L_3$ resulting from the stability analysis of the quiescent basic
state for $a\!=\!0.1$ and $L_2\!=\!2$ at $Ha=R=S=70$.  
$L_3\!=\!2\pi/k_3$ is the wavelength of the perturbation in the
$x_3$ direction (stability was tested for each wave number $k_3$ separately).
The dotted line marks the maximum growth rate of the most unstable 2D mode 
($k_3\!=\! 0$) for the same parameters.}
\label{fig9}
\end{figure}
\begin{figure}
\begin{center}
\epsfig{file=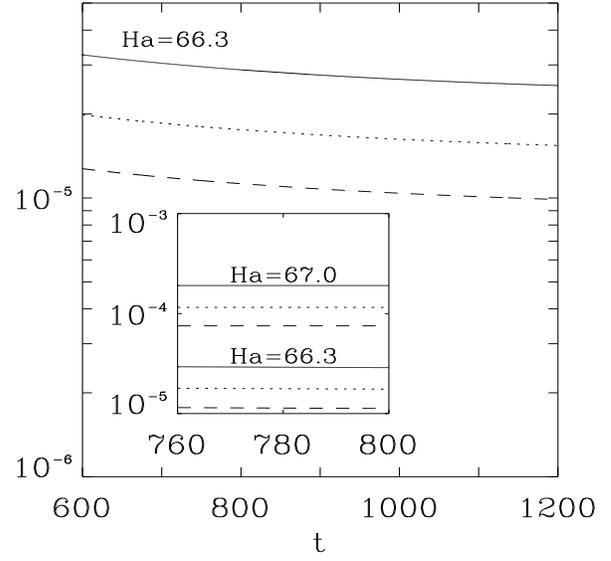,width=7cm,height=7cm}
\end{center}
\caption{Temporal behavior of the specific kinetic (dotted line), magnetic
(dashed line), and total (solid line) energies for the nonlinear two-dimensional
development of the tearing mode for $Ha\!=\!S\!=\!M\!=\!66.3\,$, $a\!=\!0.1$,
and $L_2\!=\!4$.  The inset shows additionally the corresponding development for
$Ha\!=\!S\!=\!M\!=\!67\,$.  }
\label{fig2}
\end{figure}
\begin{figure}
\begin{center}
\epsfig{file=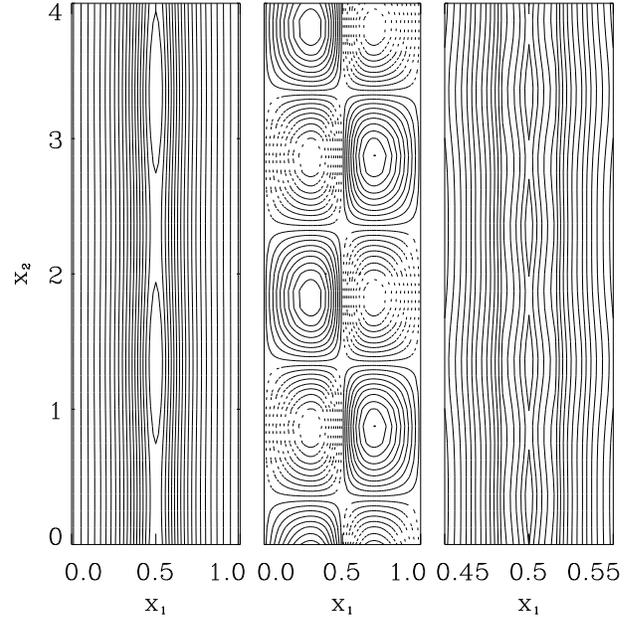,width=8cm,height=8cm}
\end{center}
\caption{Magnetic field lines (left), velocity stream lines (middle), and
contour lines of the current density component $J_3$ (right) for
$Ha\!=\!67$, $a\!=\!0.1$, and $L_2\!=\!4$ at $t\!=\!800$.  Solid (dashed)
velocity stream lines correspond to clockwise (counterclockwise) motion.
Only the inner part of the current sheet is shown for the current density 
contour plot to highlight the filamentation of $J_3$.}
\label{fig3}
\end{figure}
\begin{figure}
\begin{center}
\epsfig{file=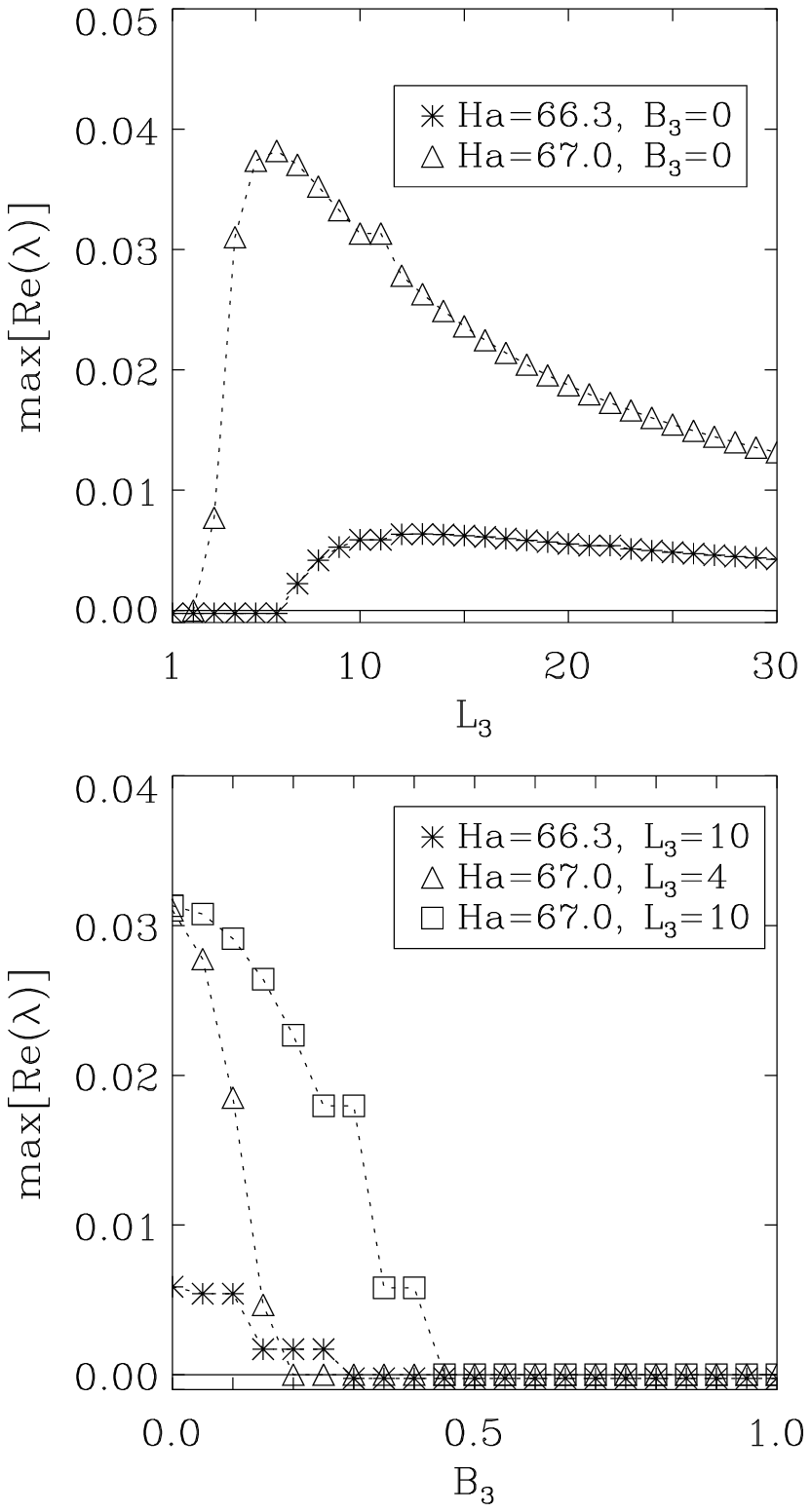,width=7cm,height=16cm}
\end{center}
\vspace{0.5cm}
\caption{Maximum real parts of the eigenvalue spectrum versus
$L_3$ (upper panel) and $\overline{B_3^e}$ (lower panel)
for $a\!=\!0.1$ and $L_2\!=\!4$ at different Hartmann numbers $Ha$ (always
$S\!=\!M$). $L_3\!=\!2\pi/k_3$ is the wavelength of the perturbation in the
$x_3$ direction (stability was tested for each wave number $k_3$ separately).}
\label{fig4}
\end{figure}
\vfill
\begin{figure}
\begin{center}
\epsfig{file=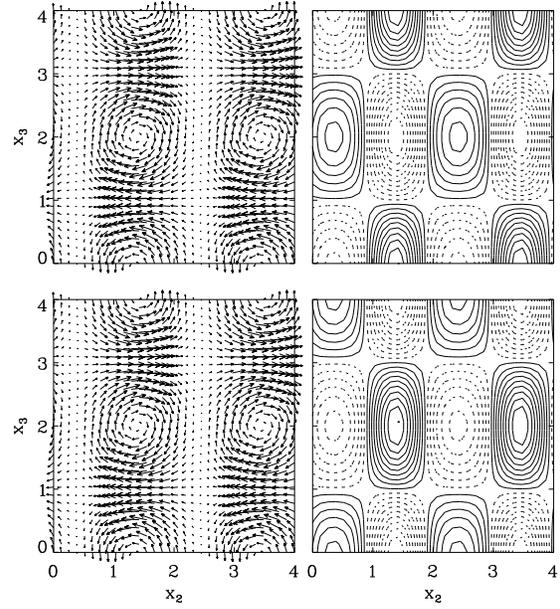,width=8cm,height=8cm}
\end{center}
\vspace{0.5cm}
\caption{Most unstable 3D eigenstate with wave number $k_3\!=\!2\pi/4$ to the
time-asymptotic 2D state (see also Fig.  \ref{fig3}) at $Ha\!=\!67$.  Vector
plots of the velocity field (left) and contour lines of the magnetic 
field component $B_3$ 
(right) in planes
$x_1\!=\! 0.6$ (upper row) and $x_1\!=\! 0.4$ (lower row) 
are shown. Dotted contour lines indicate negative $B_3$, 
solid lines positive ones.  The time-asymptotic 2D state is mixed with the 3D
perturbation in the ratio 50\% equilibrium to 50\% perturbation.}
\label{fig5}
\end{figure}
\begin{figure}
\caption{(JPG-File) Isosurfaces $|{\bf J}|\!=\!5.45$ for the same data set
as in Fig. \ref{fig5}. The maximum value of $|{\bf J}|$ is $9.77$.}
\label{fig6}
\end{figure}
\vfill
\pagebreak
\begin{figure}
\begin{center}
\epsfig{file=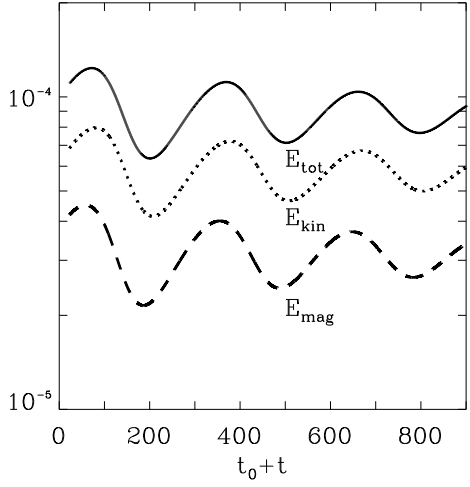,width=6cm,height=6cm}
\end{center}
\vspace{0.5cm}
\caption{Temporal behavior of the specific kinetic, magnetic, and total energies
for the nonlinear development of the three-dimensional system for
$L_2\!=\!L_3\!=\!4$, $\overline{B_2^e}\!=\!\overline{B_3^e}=0$, and
$Ha\!=\!67$. The time $t_0$ was 2900.}
\label{fig7}
\end{figure}
\vfill
\begin{figure}
\caption{(JPG-File) Isosurfaces $|{\bf v}|\!=\!0.03$ and $|{\bf v}|\!=\!0.016$ for the
time-asymptotic 2D state.  The values of the parameters are $L_2\!=\!4$,
$\overline{B_2^e}\!=\!\overline{B_3^e}=0$,
and $Ha\!=\!67$.  The maximum and minimum values of $|{\bf v}| $ are
$0.0384$ and $0.0017$, respectively.}
\label{fig8}
\end{figure}
\begin{figure}
\caption{(JPG-File) Isosurfaces $|{\bf v}|\!=\!0.03$ and $|{\bf v}|\!=\!0.016$ for the
selfconsistent 3D state at
$t\!=\!550$.  The values of the parameters are $L_2\!=\!L_3\!=\!4$, 
$\overline{B_2^e}\!=\!\overline{B_3^e}=0$,
and $Ha\!=\!67$.  The maximum and minimum values of $|{\bf v}| $ are
$0.0311$ and
$0.0$, respectively, in the 3D case.}
\label{fig8a}
\end{figure}

\end{narrowtext}
\end{multicols}
\end{document}